\documentstyle[titlepage,12pt]{article}
\begin{document}
\parindent 2em
\baselineskip 4.5ex

\noindent
June 12, 2003

\begin{center}

{\Large Theory of strongly phase fluctuating d-wave superconductors and the
spin response in underdoped cuprates}  
\vspace{15mm}

Igor F. Herbut \\ 

Physics Department, Simon Fraser University, 
Burnaby, British Columbia, Canada V5A 1S6  
\end{center}
\vspace{10mm}

\noindent
{\bf Abstract:} General theory of d-wave quasiparticles coupled to phase
fluctuations of the superconducting order parameter is discussed.
In the charge sector the superfluid density is found to conform to 
the Uemura scaling. The spin susceptibility exhibits
four distinct regimes with increasing frequency, and scales with
the superconducting $T_c$, as observed. 
\vspace{10mm}


\noindent
Fax: 1-604-291-3592, Email: iherbut@sfu.ca.

\pagebreak

The intimate relationship between antiferromagnetism and d-wave
superconductivity (dSC) has been one of the central themes in the physics
of cuprates. Recently a new connection between the
two has been put forward \cite{herbut1}, \cite{herbut2}, by
which the antiferromagnetism is the result of the dynamical breaking
of an approximate "chiral" symmetry of the standard dSC. The relevant 
interaction between the low-energy
quasiparticles is provided by the quantum vortex fluctuations, which 
are assumed to increase with underdoping.

 The low energy theory of d-wave quasiparticles and the
phase fluctuations of the superconducting order parameter at $T=0$
and in two dimensions may be written as
\cite{herbut3} $S=\int d^3 x {\cal L}$, ${\cal L}= {\cal L}_\psi +
{\cal L}_{gauge} + {\cal L}_\phi$ , where
\begin{equation}
{\cal L}_\psi = \sum_{i=1}^{2} {\bar \Psi}_i \gamma_\mu
( \partial_\mu- ia_\mu) \Psi_i ,
\end{equation}
\begin{eqnarray}
{\cal L}_{gauge} = \frac{1}{\pi} {\bf a}\cdot (\nabla\times {\bf A}_-)
+ \frac{1}{\pi} {\bf v}\cdot (\nabla\times {\bf A}_+) + \\ \nonumber
2K ({\bf v}+{\bf A})^2 + i {\bf J}_c \cdot ({\bf v}+{\bf A}),
\end{eqnarray}
\begin{equation}
{\cal L}_\phi = \sum_{n=1}^2 [ | (\nabla -i{\bf A}_n ) \Phi_n|^2
+ \mu^2 |\Phi_n|^2 + \frac{b}{2} |\Phi_n|^4 ].
\end{equation}
$\Psi_i $ describe the neutral spin-1/2
quasiparticles (spinons) near the four nodes at $\pm
{\bf K}_{1,2}$ \cite{herbut2}, complex $\Phi_{1,2}$ create the fluctuating 
vortex loops associated with spin up and down,
and ${\bf a}$, ${\bf v}$ \cite{franz} and
${\bf A}_{1,2} = {\bf A}_+ \pm {\bf A}_-$ are
the auxiliary gauge fields that facilitate the spinon-vortex coupling.
${\bf J}_c$ is the charge current, and ${\bf A}$ an electromagnetic
gauge potential. The two quasiparticle velocities have been set to
$v_F = v_\Delta=1$ in (1). $K$ is the bare superfluid
density.

When $\langle \Phi_1 \rangle =  \langle \Phi_2 \rangle  \neq 0$,
the above theory reduces to the $QED_3$ \cite{franz},
\cite{herbut2}, with the
insulating spin density wave (SDW) as the ground state \cite{babak}. 
Here I focus on the superconducting state, in which $\langle \Phi_n
\rangle =0$. Integrating out $\vec{v}$ and the
virtual vortex fluctuations yields \cite{herbut3} 
\begin{equation}
{\cal L}\rightarrow {\cal L}_\psi + \frac{6m}{\pi} {\bf a}^2
+ \frac{\rho_{sf}}{2}{\bf A}^2 + i Z {\bf J}_c \cdot
{\bf A} + O({\bf J}_c ^2),
\end{equation}
with $\rho_{sf} = 48mK/(12m+4\pi K)$. 
$Z= 12m/(12m + 4\pi K)$ is the charge renormalization factor.
$m^2 = \mu^2 +O(b)$ is the uniform vortex susceptibility,
assumed to be proportional to doping. Since on dimensional grounds
$m\sim T_c$, for $m \ll K$, $\rho_{sf} \sim T_c$, in qualitative
agreement with the Uemura scaling. Furthermore, spin may be
considered separated from the charge in Eq. (4).
Upon integration over the massive $\vec{a}$, the spin
sector becomes the 2+1 dimensional Thirring model.
In the RPA approximation the spin susceptibility becomes \cite{herbut4}
\begin{equation}
\chi '' (2{\bf K}_i\pm {\bf q},\omega )  =
(\frac{12 m}{\pi})^2  \frac{ N \Theta(\omega^2 - q^2)
\sqrt{\omega^2 - q^2 } }
{ (96 m/\pi)^2 + \omega^2 - q^2}.
\end{equation}
This describes the soft spin mode in the dSC, the condensation of which
on the insulating side yields the SDW \cite{herbut1}, \cite{herbut2}.

 Eq. (5) implies that for $\omega \ll T_c$, the maximum values of
 $\chi ''$ are located at four {\it diagonally} incommensurate
 wave vectors $\pm 2{\bf K}_{1,2}$. Due to their low intensity ($\sim
 \omega$) and narrowness, these `mother' peaks should be very hard to observe, and
 indeed have not yet been detected. As the frequency
 increases, however, both the intensity and the width of the peaks
 grow. Restoring 
 $v_F \gg v_\Delta$ in (5), and shifting the peaks to  the 
 vicinity of $(\pi,\pi)$,
 one finds the `mother' peaks overlapping first at four {\it parallel}
 incommensurate positions, independent of frequency.
 With the further increase in frequency, the initial peaks at some
 point begin also 
 to overlap at $(\pi,\pi)$, and for a while the commensurate response
 dominates. Assuming $2{\bf K}_1 = 0.465 (2 \pi,2 \pi)$ \cite{mook}, and
 $v_F = 1.2$ $eV A$ the energy of the `resonance' may be estimated to
 $\sim 60$ $meV$, not far from  $\sim 40$ $meV$ seen in slightly
 underdoped cuprates. Finally, at even larger $\omega $ the maximum in
 Eq. (5) shifts to a
 $q \neq 0$, which implies a weak redistribution of the
 commensurate peak to four parallel incommensurate positions,
 but this time with upward dispersion. 
 The evolution of the spin response \cite{herbut4} is in  
 agreement with the observations in underdoped YBCO \cite{arai}.

 This work has been supported by NSERC of Canada and the Research
 Corporation.

\end{document}